\documentclass[runningheads]{llncs}
 \pdfoutput=1 
\usepackage{graphicx}
\usepackage{longtable}
\usepackage{colortbl}
\usepackage{pdfpages}

\usepackage{multirow}
\usepackage{hyperref}
\usepackage{tcolorbox}
\usepackage{orcidlink}
\usepackage{enumitem}
\usepackage{rotating}
\usepackage{environ}
\usepackage{pdflscape}
\usepackage{stmaryrd}
\usepackage{cite}
\usepackage{amsmath,amssymb,amsfonts}
\usepackage{algorithmic}
\usepackage{textcomp}
\usepackage{xcolor}
\usepackage{placeins}
\usepackage{tabularx}
\usepackage{longtable}
\usepackage{fontawesome}
\usepackage{soul}
\usepackage{ifthen}
\usepackage{amssymb}
\newboolean{showcomments}
\setboolean{showcomments}{true} 

\ifthenelse{\boolean{showcomments}}
  {\newcommand{\nb}[2]{
    \fcolorbox{gray}{yellow}{\bfseries\sffamily\scriptsize#1}
    {\sf\small$\blacktriangleright$\textit{#2}$\blacktriangleleft$}
   }
   
  }
  {\newcommand{\nb}[2]{}
   
  }

\usepackage[normalem]{ulem} 
\usepackage{xcolor}

\ifthenelse{\boolean{showcomments}}
  {
  
  \newcommand{\del}[1]{\textcolor{red}{\sout{#1}}} 
  \newcommand{\inspar}[1]{\color{blue}{#1}\color{blue}} 
  }
  {
  
  \newcommand{\del}[1]{} 
  \newcommand{\inspar}[1]{} 
  
  }
\usepackage{caption}
\usepackage{subcaption}

\usepackage{lipsum}
\usepackage{changepage}
\usepackage{framed}
\usepackage{setspace}

\newcommand{\interviewquote}[2]{
 \def\FrameCommand{%
    \hspace{0pt}%
    {\color{blue}\vrule width 1.5pt}%
    {\color{white}\vrule width 4pt}%
    \colorbox{white}
  }%
  \MakeFramed{\advance\hsize-\width\FrameRestore}%
  \noindent\hspace{-4.55pt}
  \begin{adjustwidth}{}{7pt}
  {\scriptsize ``\textbf{#1}'' - {#2}}\vspace{0.1pt}\end{adjustwidth}\endMakeFramed%
}
\begin{document}
\title{An investigation of challenges encountered when specifying training data and runtime monitors for safety critical ML applications\thanks{This project has received funding from the European Union’s Horizon 2020 research and innovation program under grant agreement No 957197.}.\\
}
\titlerunning{Challenges when specifying data and runtime monitors}
%
\author{Hans-Martin Heyn\inst{1,2}\orcidID{0000-0002-2427-6875}
\and
Eric Knauss\inst{1,2}\orcidID{0000-0002-6631-872X} 
 \and \\
Iswarya Malleswaran\inst{1} \and
Shruthi Dinakaran\inst{1}}
\authorrunning{H.-M. Heyn et al.}
%
\institute{Chalmers University of Technology, SE-412 96 Gothenburg, Sweden \and
University of Gothenburg, SE-405 30 Gothenburg, Sweden}
%
\maketitle              

\begin{abstract}
\noindent \textbf{[Context and motivation]} 
The development and operation of critical software that contains machine learning (ML) models requires diligence and established processes. Especially the training data used during the development of ML models have major influences on the later behaviour of the system. Runtime monitors are used to provide guarantees for that behaviour. 
\noindent \textbf{[Question / problem]} We see major uncertainty in how to specify training data and runtime monitoring for critical ML models and by this specifying the final functionality of the system. In this interview-based study we investigate the underlying challenges for these difficulties.
\noindent \textbf{[Principal ideas/results]} Based on ten interviews with practitioners who develop ML models for critical applications in the automotive and telecommunication sector, we identified 17 underlying challenges in 6 challenge groups that relate to the challenge of specifying training data and runtime monitoring. 
\noindent \textbf{[Contribution]} The article provides a list of the identified underlying challenges related to the difficulties practitioners experience when specifying training data and runtime monitoring for ML models. Furthermore, interconnection between the challenges were found and based on these connections recommendation proposed to overcome the root causes for the challenges.
\end{abstract}


\section{Introduction}
\label{sec:Introduction}
With constant regularity, unexpected and undesirable behaviour of machine learning (ML) models are reported in academia \cite{Islam2019, Blodgett2020, Humbatova2020, Zhang2020, Wardat2021}, the press, and by NGOs\footnote{non-governmental organisations, e.g., \url{https://algorithmwatch.org/en/stories/}}. These problems become especially apparent, and reported upon, when ML models violate ethical principles. Racial, religious, or gender biases are introduced through a lack of insight into the (sometimes immensely large set of) training data and missing runtime checks for example in large language models such as GPT-3 \cite{Abid2021}, or facial recognition software based on deep learning \cite{Mehrabi2021}. Unfortunately, improving the performance of deep learning models often requires an exponential growth in training data \cite{Banko2001}. Data requirements can help in preventing unnecessarily large and biased datasets \cite{Vogelsang2019}. Due to changes in the environment, ML models can become ``stale", i.e., the context changes so significantly that the performance of the model decreases below acceptable levels \cite{Bayram2022}. Runtime monitors collect performance data and indicate the need for re-training of the model with updated training data. However, these monitors need to be specified at design time. Data requirements can support the specification of runtime monitor \cite{Bencomo2010}. The lack of specifications becomes specifically apparent with ML models that are part of \emph{critical} software \footnote{We define critical software as software that is safety, privacy, ethically, and/or mission critical, i.e., a failure in the software can cause significant injury or the loss of life, invasion of personal privacy, violation of human rights, and/or significant economic or environmental consequences \cite{Knight2002}.} because it is not possible to establish traceability from system requirements (e.g., functional safety requirements) to requirements set on the training data and the runtime monitoring \cite{Marques2019}.


\vspace{-10pt}
\begin{figure}[!htb]
    \centering
    \includegraphics[width=0.9\textwidth]{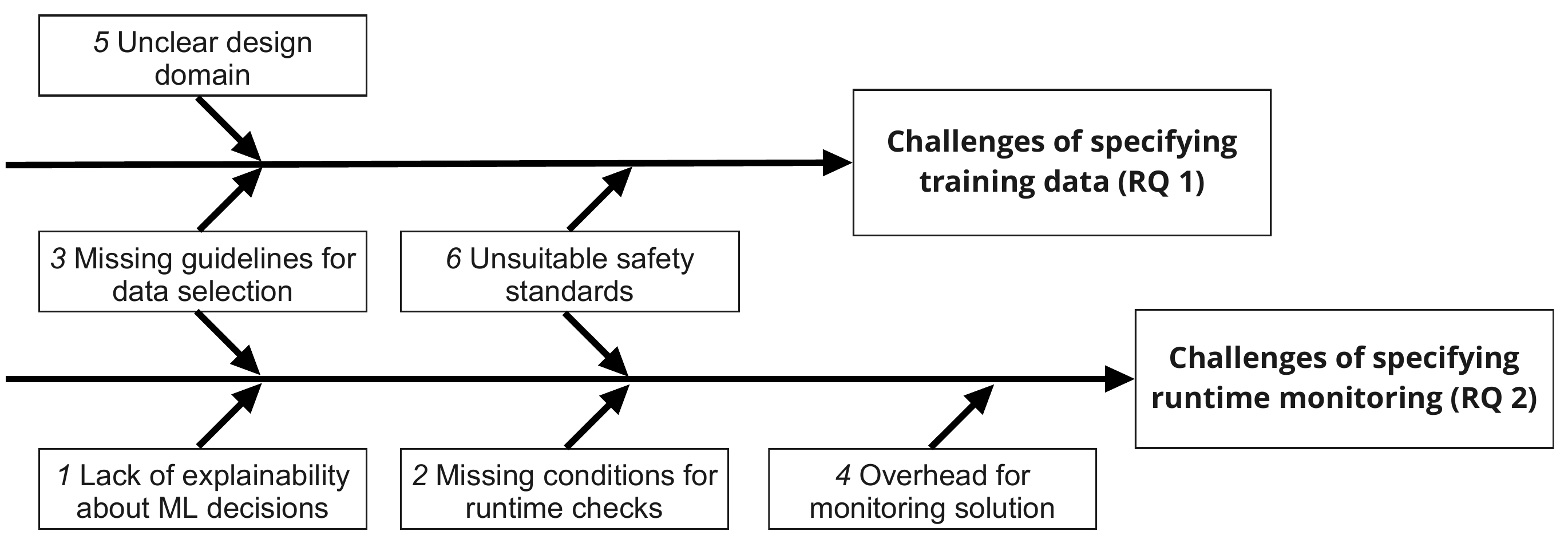}
    \caption{Overview of identified challenge categories}
    \label{fig:Cause-Effect-all}
\end{figure}
\vspace{-25pt}

\subsection*{Scope and research questions}
\vspace{-5pt}
\label{subsec:ResearchQuestions}
The purpose of this study is to highlight current challenges experienced by practitioners in specifying training data and runtime monitoring for ML in safety critical software. 

The paper contributes a practitioner's point of view on the challenges reported in academic literature. 
The aim is to identify starting-points for a future engineering research on the use of runtime monitors for critical ML systems. The following research questions guided this study:

\vspace{-5pt}
\begin{description}
\item[RQ1:] What are challenges encountered by practitioners when specifying training data for ML models in safety critical software?
\item[RQ2:] What are challenges encountered by practitioners when specifying runtime monitors especially in relation to fulfilling safety requirements?
\end{description}
\vspace{-5pt}

Figure~\ref{fig:Cause-Effect-all} shows the main themes we found in answering the research questions. Concerning RQ1, the interviewees reported on several problems: the data selection process is nontransparent and guidelines especially towards defining suitable measures for data variety are missing. There are no clear context definitions that help in defining data needs, and current safety standards provide little guidance. Concerning RQ2, we found that the problem of defining suitable metrics and the lack of guidance from safety standards also inhibits the ability to specify runtime monitors. Furthermore, practitioners reported on challenges regarding explainability of ML decisions, and the processing and memory overhead caused by runtime monitors in safety critical embedded systems. \par

The remaining sections of this paper are structured as follows: Section 2 outlines and argues for the research methods of this study; Section 3 presents the results amd answers to the research questions; Section 4 discusses the findings, provides recommendations to practitioners and for further research, identifies related literature, elaborates on threats to validity, and provides a conclusion.


\section{Research Method} \label{sec:method}
\label{sec:ResearchMethod}
We applied a qualitative interview-based survey with open-ended semi-structured interviews for data collection. 
Following the suggestions of Creswell and Creswell \cite{Creswell2017} the qualitative study was conducted in four steps: Preparation of interviews, data collection through interviews, data analysis, and result validation. 
\vspace{-5pt}
\subsubsection{Preparations of interviews}
Based on the a-priori formulated research questions, two of the researchers of this study created an interview guide\footnote{The interview guide is available at  \url{https://doi.org/10.7910/DVN/WJ8TKY}.} which was validated and improved by the remaining two researchers. The interview guide contains four sections of questions: the first section includes questions about the interviewees' current role, background and previous experiences. The second section focuses on questions that try to understand challenges when specifying and selecting training data for ML models and how training data affect the performance of these models. The third section investigates challenges when ML models are incorporated in critical systems and how they affect the ability to specify training data. The fourth section concentrates on the run time monitoring aspect of the ML model and contains questions on challenges when specifying runtime monitors.
\vspace{-5pt}
\paragraph{Sampling strategy:} We chose the participants for this study purposefully using a maximum variation strategy \cite{Creswell2017b}. We were able to recruit interviewees from five different companies, ranging from a local start-up to a multinational world leading communication company. An overview is given in Table~\ref{tab:companies}. 

\begin{table*}[!tbp]
\centering
\scriptsize
\caption{Companies participating in the study}
\label{tab:companies}
\begin{tabular}{clll}
\hline
\multicolumn{1}{c}{\textbf{\begin{tabular}[c]{@{}c@{}}Company\end{tabular}}} & \multicolumn{1}{c}{\textbf{\begin{tabular}[c]{@{}c@{}}Area of operations\end{tabular}}} & \multicolumn{1}{c}{\textbf{Employees}}& \multicolumn{1}{c}{\textbf{Countries}} \\ \hline
\rowcolor[HTML]{EFEFEF} 
1 & \begin{tabular}[l]{@{}l@{}}Telecommunication networks \end{tabular}&  $>$ 10.000 & World  \\
2 & Automotive OEM &  $>$ 10.000 & World \\
\rowcolor[HTML]{EFEFEF} 
3 & Automatic Driving &  $>$ 1.000 & Europe \\ 
4 & \begin{tabular}[l]{@{}l@{}}Industrial camera systems\end{tabular} & $>$ 1000 & USA \\
\rowcolor[HTML]{EFEFEF} 
5 & \begin{tabular}[l]{@{}l@{}}Deep Learning optimisation for IoT \end{tabular} & $>$ 100 & Sweden \\
\hline
\end{tabular}
\vspace{-10pt}
\end{table*}

\noindent A selection criteria for the company was that they must work with safety-critical systems and ML. Within the companies we tried to find interview candidates with different roles and work experiences to obtain a view beyond the developers' perspective. Besides function developers and ML model developers, we were interested in interviewing requirement engineers and product / function owners because they represent key roles in deriving system or function specifications. We provided the companies with a list of roles that we identified beforehand as interesting for interviewing\footnote{The list included functional safety experts, requirement engineers, product owners or function owners, function or model developers, and data engineers.}. Additionally, we interviewed two researchers from academia who participate in a joint industry EU Horizon 2020 project called VEDLIoT\footnote{Very efficient deep learning in the Internet of Things}. Both researchers worked also with ML models in industry before. Therefore, they could provide insights into both the academic and the industry perspective. A list of the ten interviewees for this study is provided in Table~\ref{tab:participants}.
\begin{table*}[!t]
\centering
\scriptsize
\caption{Participants of the study}
\label{tab:participants}
\begin{tabular}{cll}
\hline
\multicolumn{1}{c}{\textbf{\begin{tabular}[c]{@{}c@{}}Inter-\\ viewee\end{tabular}}} & \multicolumn{1}{c}{\textbf{Role}} & \multicolumn{1}{c}{\textbf{Experience}} \\ \hline
\rowcolor[HTML]{EFEFEF} 
A & Researcher (Academic) &  Functional Safety for ADAS \\
B & Function developer &  Sensor and perception systems \\
\rowcolor[HTML]{EFEFEF} 
C & Principal engineer &  ML model integration \\ 
D & ML model developer & \begin{tabular}[l]{@{}l@{}}Distributed and edge systems\end{tabular} \\
\rowcolor[HTML]{EFEFEF} 
E & Function owner & ADAS perception functions \\
F & \begin{tabular}[l]{@{}l@{}} Function developers \\ and test engineer \end{tabular} & Automatic driving systems\\
\rowcolor[HTML]{EFEFEF} 
G & Data Scientist & Distributed systems \\
H & Requirement Engineer & Perception systems \\
\rowcolor[HTML]{EFEFEF} 
I & Researcher (Academic) & \begin{tabular}[l]{@{}l@{}} Neural Network development \end{tabular} \\
J & Functional Safety Manager & Sensor systems \\ \hline
\multicolumn{3}{l}{\begin{scriptsize}ADAS: Advanced Driver Assistance Systems \end{scriptsize}}
\end{tabular}
\vspace{-10pt}
\end{table*}
\vspace{-5pt}
\subsubsection{Data collection through interviews}
All interviews were conducted remotely using either the conference software Zoom or Microsoft Teams and took between 60 - 90 minutes. The a-priori defined interview guide was only available to the interviewers and was not distributed to the participants beforehand. Each participant was interviewed by two interviewers who alternated in asking questions and observing. At the start of each interview, the interviewers provided some background information about the study's purpose. Then, the interview guide was followed. However, as we encouraged discussions with the interviewees, we allowed deviations from the interview guide by asking additional questions, or changing the order of the questions when it was appropriate \cite{King2018}. All interviews were recorded and semi-automatically transcribed. The interviewers manually checked and anonymised the results. \par
\vspace{-5pt}
\subsubsection{Data analysis}
The data analysis followed suggestions by Saldana \cite{Saldana2013} and consisted of two cycles of coding and validation of the themes through a workshop and member checking. 
\vspace{-10pt}
\paragraph{First coding cycle:} Attribute coding was used to extract information about the participants' role and previous experiences. Afterwards, the two interviewers independently applied structural coding to collect phrases in the interviews that represent topics relevant to answering the research questions. The researchers compared the individually assigned codes and applied descriptive coding with the aim of identifying phrases that describe common themes across the interviews.
\vspace{-10pt}
\paragraph{Theme validation:} In a focus group, the identified themes were presented and discussed. Thirteen researchers from both industry and academia in the VEDLIoT project participated. Three of the participants also were interviewed for this study. The aim of the focus group was to reduce bias in the selection of themes and to identify any additional themes that the researchers might have missed.
\vspace{-10pt}
\paragraph{Second coding cycle:} After the themes were identified and validated, the second coding cycle was used to map the statements of the interviewees to the themes, and consequently identify the answers to the research questions. The second cycle was conducted by the two researchers who did not conduct the first cycle coding in order to reduce confirmation bias. The mapping was then confirmed and agreed upon by all involved researchers.
\vspace{-10pt}
\subsubsection{Result validation}
Member checking, as described in \cite[Ch. 9]{Creswell2017b} was used to validate the identified themes that answer RQ~1 and RQ~2. Additionally, we presented the results in a 60 minute focus group to an industry partner and allowed for feedback and comments on the conclusions we drew from the data.


\section{Results} \label{sec:results}
During the first coding cycle, structural coding resulted in 117 statements for RQ1 and 77 statements for RQ2. Through descriptive coding preliminary themes were found. The statements and preliminary themes were discussed during a workshop. Based on the feedback from the workshop, 117 statements for RQ1 were categorised into eight final challenge themes and three challenge categories relating to the challenge of specifying training data. Similar, the 77 original statements for RQ2 were categorised into 13 final challenge themes in five challenge categories relating to the challenge of specifying runtime monitoring. A total of six challenge categories emerged for both RQs, out of which two categories contain challenges relating to both training data and runtime monitoring specification, and three challenge themes base on statements from both RQs. The categories and final challenge themes are listed in Table~\ref{tab:themes}.

\begin{table*}[!t]
\centering
\scriptsize
\caption{Challenge groups (bold) and themes found in the interview data. Data.: Challenges related to specifying training data (RQ1). Monitor.: Challenges related to specifying runtime monitoring (RQ2).}
\label{tab:themes}
\begin{tabular}{clccc}
\hline
\multicolumn{1}{c}{}                              & \multicolumn{1}{c}{}                                           & \multicolumn{2}{c}{\textbf{Relates to}} & \multicolumn{1}{c}{\textbf{Related}}                               \\
\multicolumn{1}{c}{\multirow{-2}{*}{\textbf{ID}}} & \multicolumn{1}{c}{\multirow{-2}{*}{\textbf{Challenge Theme}}} & \multicolumn{1}{c}{\textbf{Data.}} & \multicolumn{1}{c}{\textbf{Monitor.}} & \multicolumn{1}{c}{\textbf{Literature}} \\ \hline
\rowcolor[HTML]{EFEFEF} 
\textbf{1}                                        & \textbf{Lack of explainability about ML decisions}             & \textbf{}                         & \textbf{\checkmark} &                        \\
\rowcolor[HTML]{EFEFEF} 
1.1                                               & No access to inner states of ML models                         &                               & \checkmark  &  [18]                            \\
\rowcolor[HTML]{EFEFEF} 
1.2                                               & No failure models for ML models                                &                                   & \checkmark &    [51]                             \\
\rowcolor[HTML]{EFEFEF} 
1.3                                               & IP protection                                                  &                                   & \checkmark &                                \\
\textbf{2}                                       & \textbf{Missing conditions for runtime checks}                 & \textbf{}                         & \textbf{\checkmark} &                       \\
2.1                                              & Unclear metrics and/or boundary conditions          &                                   & \checkmark  & [11,21,43]                              \\
2.2                                              & Unclear measure of confidence                                  &                                   & \checkmark & [17,34]                                \\
\rowcolor[HTML]{EFEFEF} 
\textbf{3}                                      & \textbf{Missing guidelines for data selection}                 & \textbf{\checkmark}                        & \textbf{\checkmark}   &                     \\
\rowcolor[HTML]{EFEFEF} 
3.1                                             & Disconnection from requirements                                & \checkmark                                 &      & [16,42]                        \\
\rowcolor[HTML]{EFEFEF} 
3.2                                             & Grown data selection habits                                    & \checkmark                                 &                 &  [20,33]                \\
\rowcolor[HTML]{EFEFEF} 
3.3                                             & Unclear completeness criteria                                   & \checkmark                                 &              &  [49]                   \\
\rowcolor[HTML]{EFEFEF} 
3.4                                             & Unclear measure of variety                                     & \checkmark                                 & \checkmark     &   [45,50]                         \\
\textbf{4}                                       & \textbf{Overhead for monitoring solution}                      & \textbf{}                         & \textbf{\checkmark}   &                     \\
4.1                                              & Limited resources in embedded systems                          &                                   & \checkmark                &  [38]               \\
4.2                                              & Meeting timing requirements                                    &                                   & \checkmark                &                 \\
4.3                                              & Reduction of true positive rate                                &                                   & \checkmark              &                   \\
\rowcolor[HTML]{EFEFEF} 
\textbf{5}                                        & \textbf{Unclear design domain}                                 & \textbf{\checkmark}                        & \textbf{}           &              \\
\rowcolor[HTML]{EFEFEF} 
5.1                                               & Design domain depends on available data                        & \checkmark                                 &             & [6]                     \\
\rowcolor[HTML]{EFEFEF} 
5.2                                               & Uncertainty in context                                         & \checkmark                                 &            & [22]                    \\
\textbf{6}                                       & \textbf{Unsuitable safety standards}                           & \textbf{\checkmark}                        & \textbf{\checkmark}    &                    \\
6.1                                              & Focus on processes instead of technical solution            & \checkmark                                 &  \checkmark     &  [10]                          \\
6.2                                              & No guidelines for probabilistic effects in software                   &  \checkmark                                 &        &  [28,43]                        \\
6.3                                              & Safety case only through monitoring solution      &                                  &    \checkmark               & [31,46]              \\ \hline
\end{tabular}
\vspace{-10pt}
\end{table*}

\vspace{-10pt}
\subsection{Answer to RQ1: Challenges practitioners experience when specifying training data}
\vspace{-5pt}
The interviewees were asked to share their experiences in selecting training data, the influence of the selection of training data on the system's performance and safety, and any experiences and thoughts on defining specifications for training data for ML. Based on the interview data, we identified three challenge groups related to specifying training data: missing guidelines for data selection, unclear design domain, and unsuitable safety standards

\vspace{-10pt}
\subsubsection{Missing guidelines for data selection}
Four interviewees reported on a lack of guidelines and processes related to the selection of training data. A reason can be that data selection bases on ``grown habits" that are not properly documented. Unlike conventional software development, the training of ML is an iterative process of discovering the necessary training data based on experience and experimentation.
Requirements set on the data are described as disconnected and unclear for the data selection process. For example, one interviewee stated that if a requirements is set that images shall contain a road, it remains unclear what specific properties this road should have. Six interviewees described missing requirements on the data variety and missing completeness criteria as a reason for the disconnection of requirements from data selection. 

\interviewquote{How much of it (the data) should be in darkness? How much in rainy conditions, and how much should be in snowy situations?}{Interview~F}
\vspace{-20pt}
\interviewquote{For example, we said that we shall collect data under varying weather conditions. What does that mean?}{Interview~B}
\vspace{-5pt}
\noindent Another interviewee stated that it is not clear how to measure variety, which could be a reason why it is difficult to define requirements on data variety.
\vspace{-5pt}
\interviewquote{What [is] include[d] in variety of data? Is there a good measure of variety?}{Interview~A}


\vspace{-20pt}
\subsubsection{Unclear design domain}
Three interviewees describe uncertainty in the design domain as a reason for why it is difficult to specify training data. If the design domain is unclear, it will be challenging to specify the necessary training data. 

\interviewquote{We need to understand for what context the training data can be used.}{Interview~J}
\vspace{-18pt}
\interviewquote{ODD [(Operational Design Domain)]? Yes, of course it translates into data requirements.}{Interview~F}

\vspace{-20pt}
\subsubsection{Unsuitable safety standards}
Because we were specifically investigating ML in safety critical applications, we asked the participants if they find guidance in safety standards towards specifying training data. Five interviewees stated that current safety standards used in their companies do not provide suitable guidance for the development of ML models. While for example ISO~26262 provides guidance on how to handle probabilistic effects in hardware, no such guidance is provided for software related probabilistic faults. 

\interviewquote{The ISO~26262 gives guidance on the hardware design; [...] how many faults per hour [are acceptable] and how you achieve that. For the software side, it doesn't give any failure rates or anything like that. It takes a completely process oriented approach [...].}{Interview~C}

\noindent One interviewee mentioned that safety standards should emphasise more the data selection to prevent faults in the ML model due to insufficient training.
\vspace{-5pt}
\interviewquote{To understand that you have the right data and that the data is representative, ISO~26262 is not covering that right now which is a challenge.}{Interview~H}

\vspace{-10pt}
\subsection{Answer to RQ2: Challenges practitioners experience when specifying runtime monitors}
\vspace{-5pt}
We asked the interviewees on the role of runtime monitoring for the systems they develop, their experience with specifying runtime monitoring, and the relation of runtime monitoring to fulfilling safety requirements on the system.
We identified five challenge groups related to runtime monitoring: lack of explainability about ML decisions, missing conditions for runtime checks, missing guidelines for data selection, overhead for monitoring solution, and unsuitable safety standards.

\vspace{-10pt}
\subsubsection{Lack of explainability about ML}
A reason why it is difficult to specify runtime monitors for ML models is the inability to produce failure models for ML. In normal software development, causal cascades describe how a fault in a software components propagates trough the systems and eventually leads to a failure. This requires the ability to break down the ML model into smaller components and analyse their potential failure behaviour. Four interviewees however reported that they can only see the ML model as a ``black-box" with no access to the inner states of the ML model. As a consequence, there is no insight into the failure behaviour of the ML model. 

\interviewquote{[Our insight is] limited because it's a black box. We can only see what goes in and then what comes out to the other side. And so if there is some error in the behavior, then we don't know if it's because [of a] classification error, planning error, execution error?}{Interview~F}

\noindent The reason for opacity of ML models is not necessarily due to technology limitations, but also due to constraints from protection of intellectual property (IP).

\interviewquote{Why is it a black box? That's not our choice. That's because we have a supplier and they don't want to tell us [all details].}{Interview~F}

\vspace{-15pt}
\subsubsection{Missing conditions for runtime checks}
A problem of specifying runtime monitors is the need for finding suitable monitoring conditions. This requires the definition of metrics, goals and boundary conditions. Five interviewees report that they face challenges when defining these metrics for ML models.

\interviewquote{What is like a confidence score, accuracy score, something like that? Which score do you need to ensure [that you] classified [correctly]?}{Interview~F}

\noindent Especially the relation between correct behaviour of the ML model and measure of confidence is unclear, and therefore impede runtime monitoring specification.

\interviewquote{We say confidence, that's really important. But what can actually go wrong here?}{Interview~J}

\vspace{-15pt}
\subsubsection{Missing guidelines for data selection}
The inability to specify the meaning of data variety also relates to missing conditions for runtime checks. For example, runtime monitors can be used to collect additional training data, but without a measure of data variety it is difficult to find the required data points. 

\vspace{-10pt}
\subsubsection{Overhead for monitoring solution}
An often overlooked problem seems to be the induced (processing) overhead from a monitoring solution. Especially in the automotive sector, many software components run on embedded computer devices with limited resources. 

\interviewquote{You don't have that much compute power in the car, so the monitoring needs to be very light in its memory and compute footprint on the device, maybe even a separate device that sits next to the device.}{Interview~F}

\noindent And due to the limited resources in embedded systems, monitoring solutions can compromise timing requirements of the system. Additionally, one interviewee reported concerns regarding the reduction of the ML model's performance.

\interviewquote{[…] the true positive rate is actually decreasing when you have to pass it through this second opinion goal. It's good from a coverage and safety point of view, but it reduces the overall system performance.}{Interview~F}

\vspace{-15pt}
\subsubsection{Unsuitable safety standards}
Safety standards are mostly not suitable for being applied to ML model development. Therefore, safety is often ensured through non-ML monitoring solutions. Interviewees reported that it is not a good solution to rely only on the monitors for safety criticality: 

\interviewquote{[…] so the safety is now moved from the model to the monitor instead, and it shouldn't be. It should be the combination of the two that makes up safety.}{Interview~B}

\noindent One reason is that freedom of inference between a non-safety critical component (the ML model), and a safety critical component (the monitor) must be ensured which can complicate the system design.

\interviewquote{And then especially if you have mixed critical systems [it] means you have ASIL [(Automotive Safety Integrity Level)] and QM [(Quality Management)] components in your design [and] you want to achieve freedom from interference in your system. You have to think about safe communication and memory protection.}{Interview~J}

\section{Discussion and Conclusion} \label{sec:discussion}
The results reveal connections between the challenges. 
Not all theme groups relate exclusively to one of the two challenges. 
For example, themes in the groups \emph{unsuitable safety standards} and \emph{missing guidelines for data selection} relate to both challenges of specifying training data and runtime monitoring. 
Furthermore, we identified cause-effect relations between different themes and across different group of themes.
For example \emph{IP protection} is a cause for the \emph{inability of accessing the inner states} and for \emph{creating failure models for ML model}. We based this assessment on a semantic analyses of the words used in the statements relating to these themes. For example, Interviewee~F stated that:
\vspace{-5pt}
\interviewquote{That neural network is something [of a] black box in itself. You don't know why it do[es] things. Well, you cannot say anything about its inner behavior}{Interview~F} 
\vspace{-5pt}
\noindent Later in the interview, the same participants states: 
\vspace{-5pt}
\interviewquote{Why is it a black box? That's not our choice.  That's because we have a supplier and they don't want to tell us [all details].}{Interview~F}
\vspace{-5pt}

\begin{figure}[!h]
    \centering
    \includegraphics[width=0.8\textwidth]{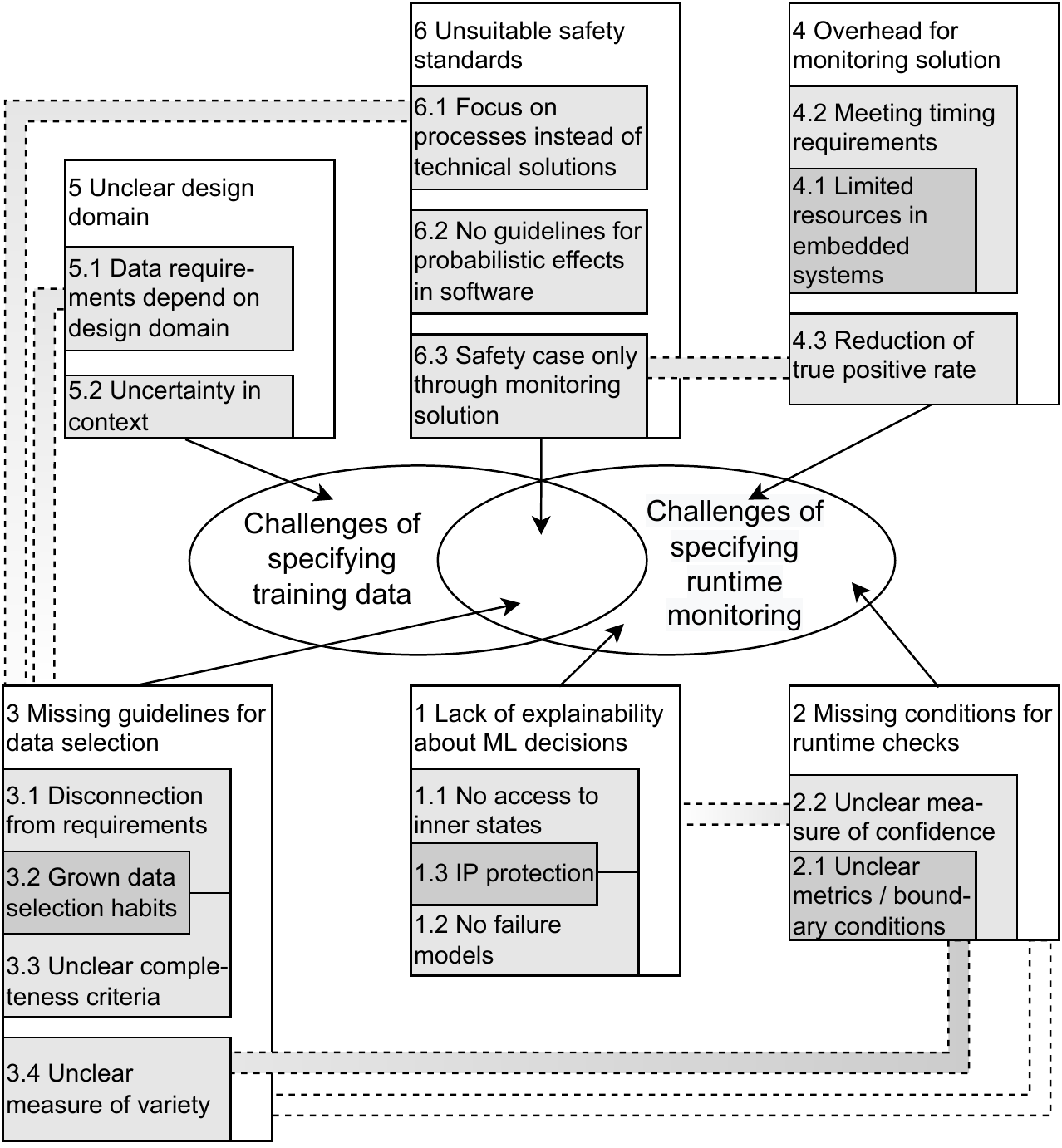}
    \caption{Connection between the identified challenge themes. Enclosed themes have been identified as causes for the surrounding themes. Furthermore, dotted lines indicate relations between different themes.}
    \label{fig:Challenges_connection}
\vspace{-10pt}
\end{figure}

\noindent Figure~\ref{fig:Challenges_connection} illustrates the identified cause-effect relations, relations between the themes, and how the different themes relate to the challenges.

\vspace{-10pt}
\subsubsection{Recommendations to practitioners and for further research}
The identified root causes of the challenges described by the participants allowed us to formulate recommendations listed in Table~\ref{tab:recommendation}. Because these recommendations try to solve root causes described by the participants of the interview study, we think they are a useful first step towards solving the challenges related to specifying training data and runtime monitoring.


\begin{table}[!t]
\centering
\scriptsize
\caption{Recommendations for practitioners and suggestions for further research}
\label{tab:recommendation}
\begin{tabular}{cp{11.5cm}}
\hline
\multicolumn{1}{c}{\textbf{ID}} &
  \multicolumn{1}{c}{\textbf{Recommendation}} \\ \hline
\textbf{I} &
  \textbf{Avoid restrictive IP protection.} IP protection is a cause for the inability of accessing the inner states of the ML models (black-box model). This causes a nontransparent measure of confidence, and an inability to formulate failure models. To our knowledge, no studies have yet been performed on the consequences of IP protection of ML models on the ability to monitor and reason (e.g., in a safety case) for the correctness of ML model decisions. \\ \hline
\rowcolor[HTML]{EFEFEF} 
\textbf{II} &
  \textbf{Relate measures of confidence to actual performance metrics.} For runtime monitoring, the measure of confidence is often used to evaluate the reliability of the ML model's results. But without understanding and relating that measure to clearly defined performance metrics of the ML model first, the measure of confidence provides little insight for runtime monitoring. In general, defining suitable metrics and boundary conditions should become an integral part of RE for machine learning as it affects both the ability to define data requirements and runtime monitoring requirements. \\ \hline
\textbf{III} &
  \textbf{Overcome grown data selection habits.} Grown data selection habits have been mentioned as a reason for a lack of clear completeness criteria and a disconnection from requirements. Based on our results, we argue that more systematic data selection processes need to be established in companies. This would allow for a better connection of the data selection process to requirement engineering and it creates a traceability between system requirements, completeness criteria and data requirements. Additionally, it might also reduce the amount of data needed for training, and therefore cost of development. \\ \hline
\rowcolor[HTML]{EFEFEF} 
\textbf{IV} &
  \textbf{Balance hardware limitation in embedded systems.} Runtime monitoring causes a processing and memory overhead that can compromise timing requirements and reduce the ML model's performance. Today, safety criticality of systems with ML is mostly ensured through monitoring solutions. By decomposing the safety requirements instead onto both the monitoring and the ML model, the monitors might become more resource efficient, faster, and less constraining in regards to the decisions of the ML model. However, safety requirements on the ML models might trigger requirements on the training data. \\ \hline
\end{tabular}
\end{table}

\subsection{Related Literature}
\vspace{-5pt}
\paragraph{The problem of finding the ``right" data:} For acquiring data, data scientists have to rely on data mining with little to no quality checking and potential biases \cite{Barocas2016}. Biased datasets are a common cause for erroneous or unexpected behaviour of ML models in critical environments, such as in medical diagnostic \cite{Bernhardt2022}, in the juridical system \cite{Goodman2017, Miron2021}, or in safety-critical applications \cite{Uricar2019, Fabbrizzi2021}. \par
There are attempts to create ``unbiased" datasets. One approach is to curate manually the dataset, such as in the FairFace dataset \cite{Karkkainen2021}, the CASIA-SURF CeFaA dataset \cite{Liu2021}, or Fairbatch \cite{Roh2021}. An alternative road is to use data augmentation techniques to ``rebalance" the dataset \cite{Uchoa2020, Jaipuria2020}. However, it was discovered that it is not sufficient for avoiding bias to use an assumed balanced datasets during training \cite{Wang2019, Gwilliam2021, Wang2022} because it is often unclear which features in the data need to be balanced. Approaches for curating or manipulating the dataset require information on the target domain, i.e., one needs to set requirements on the dataset depending on the desired operational context \cite{Fauri2017, Bencomo2021, Heyn2022}. But deriving a data specification for ML is not common practise \cite{Ishikawa2019, Liu2020, Sambasivan2021}. 

\paragraph{The problem of finding the ``right" runtime monitor:} Through clever test strategies, some uncertainty can be eliminated in regards to the behaviour of the model \cite{Breck2017}. However, ML components are often part of systems of systems and their behaviour is hard to predict and analyse at design time \cite{Vierhauser2014}. DevOps principles from software engineering give promising ideas on how to tackle remaining uncertainty at runtime \cite{Lwakatare2020}. As part of the operation of the model, runtime models that ``augment information available at design-time with information monitored at runtime" help in detecting deviations from the expected behaviour \cite{Giese2014}. These runtime models for ML can take the form of model assertions, i.e., checking of explicitly defined attributes of the model at runtime \cite{Kang2020}. However, the authors state that ``bias in training sets are out of scope for model assertion". Another model based approach can be the creation of neuron activation patterns for runtime monitoring \cite{Cheng2019}. Other approaches treat the ML model as ``black-box", and only check for anomalies and drifts in the input data \cite{Rahman2021} the output \cite{Shao2020}, or both \cite{Ginart2022}. However, similar to the aforementioned challenges when specifying data for ML, runtime monitoring needs an understanding on how to ``define, refine, and measure quality of ML solutions" \cite{Horkoff2019}, i.e., in relation to non-functional requirements one needs to understand which quality aspects are relevant, and how to measure them \cite{Habibullah2021}. Most commonly applied safety standards emphasise processes and traceability to mitigate systematic mistakes during the development of critical systems. Therefore, if the training data and runtime monitoring cannot be specified, a traceability between safety goals and the deployed system cannot be established \cite{Borg2018}. \par
For many researchers and practitioners, runtime verification and monitoring is a promising road to assuring safety and robustness for ML in critical software \cite{Breck2017, Ashmore2021}. However, runtime monitoring also creates a processing and memory overhead that needs to be considered especially in resource-limited environments such as embedded devices \cite{Rabiser2019}. \par
The related work has been mapped to the challenges identified in the interview study in Table~\ref{tab:themes}.

\vspace{-5pt}
\subsection{Threats to validity}
\vspace{-5pt}
A lack of rigour (i.e., degree of control) in the study design can cause confounding which can manifest bias in the results \cite{Slack2001}. The following mechanisms in this study tried to reduce confounding: The interview guide was peer-reviewed by an independent researcher, and a test session of the interview was conducted. To reduce personal bias, at least two authors were present during all interviews, and the authors took turn in leading the interviews. To confirm the initial findings from the interview study and reduce the risk of researchers' bias, a workshop was organised which was also visited by participants who were not part of the interview study. Another potential bias can arise from the sampling of participants. Although we applied purposeful sampling, we still had to rely on the contact persons of the companies to provide us with suitable interview candidates. We could not directly see a list of employees and choose the candidates ourselves. Regarding generalisability of the findings, the limited number of companies involved in the study can pose a threat to external validity. However, two of the companies are world-leading companies in their fields, which, in our opinion, gives them a deep understanding and experience of the discussed problems. Furthermore, we included companies from a variety of different fields to establish better generalisability. Furthermore, our data includes only results valid for the development of safety-critical ML models. We assume that the findings are applicable also to other forms of criticality, such as privacy-critical, but we cannot conclude on that generalisability based on the available data.




\subsection{Conclusion} \label{sec:conclusion}
This paper reported on a interview-based study that identified challenges related to specifying training data needs and runtime monitoring for safety critical ML models. Through interviews conducted at five companies we identified 17 challenges in six groups. Furthermore, we performed a semantic analysis to identify the underlying root-causes. We saw that several underlying challenges affect both the ability to specify training data and runtime monitoring. For example, we concluded that restrictive IP protection can cause an inability to access and understand the inner states of a ML model. Without insight into the ML model's state, the measure of confidence cannot be related to actual performance metrics. Without clear performance metrics, it is difficult to define the necessary degree of variety in the training data. Furthermore, grown data selection impedes proper requirement engineering for training data. Finally, safety requirements should be distributed on both the ML model which can cause requirements on the training data, and on runtime monitors which can reduce the overhead by the monitoring solution. These recommendations will serve as starting point for further engineering research.


%
%
%
\bibliographystyle{splncs04}
\bibliography{bib}

\end{document}